\documentclass[a4paper,12pt]{article}

\usepackage{amsmath}
\usepackage{amsthm}
\usepackage{amssymb}
  \newtheorem{df}{Definition}[section]
  \newtheorem{thm}[df]{Theorem}
  \newtheorem{lem}[df]{Lemma}
  \newtheorem{rmk}[df]{Remark}
  \newtheorem{prop}[df]{Proposition}
  \newtheorem{cor}[df]{Corollary}
\newcommand{\adag}{a^{\dag{}}}
\newcommand{\xnU}{x^{\underline{n}}}

\newcommand{\xnO}{(x+\epsilon)^{\overline{n}}}

\newcommand{\NnU}{N^{\underline{n}}}
\newcommand{\NnkU}{N^{\underline{n-k}}}
\newcommand{\NnO}{(N+\epsilon)^{\overline{n}}}
\newcommand{\NnkO}{(N+\epsilon)^{\overline{n-k}}}
  \makeatletter
   
   \@addtoreset{equation}{section}
  \makeatother
\begin{document}
 \title{ A New Symmetric Expression of Weyl Ordering }
 \author{ Kazuyuki FUJII\thanks{Department of Mathematical Sciences, 
  Yokohama City University, 
  Yokohama 236-0027, 
  Japan, \endgraf 
  {\it E-mail address}: fujii{\char'100}yokohama-cu.ac.jp} \ and 
  Tatsuo SUZUKI\thanks{Department of Mathematical Sciences, 
  Waseda University, 
  Tokyo 169-8555, 
  Japan, \endgraf 
  {\it E-mail address}: suzukita{\char'100}gm.math.waseda.ac.jp}}
 \date{}
 \maketitle
\begin{abstract}
For the creation operator $\adag $ and the annihilation 
operator $a$ of a harmonic oscillator, 
we consider Weyl ordering expression of $(\adag a)^n$ 
and obtain a new symmetric expression of Weyl ordering 
w.r.t. $\adag a \equiv N$ and $a\adag =N+1$ where $N$ is the number operator. 
Moreover, we interpret intertwining formulas of various orderings in view of 
the difference theory. Then we find that the noncommutative parameter 
corresponds to the increment of the difference operator w.r.t. variable $N$. 
Therefore, quantum (noncommutative) calculations of harmonic oscillators 
are done by classical (commutative) ones of the number operator 
by using the difference theory. 
As a by-product, nontrivial relations including the Stirling number 
of the first kind are also obtained.
\end{abstract}

\section{Introduction}
The problem for ordering of operators in Quantum Physics has been 
investigated then and now, 
see, for example, \cite{G}, \cite{HH}, \cite{K}, \cite{BPS} 
and their references. 

Quantization 
$$ \{ f,g \}_{\mbox{\tiny{Pb}}} \rightarrow 
 -\frac{i}{\hbar} [\hat{f}, \hat{g}] $$
is a Lie algebra homomorphism, but not a ring homomorphism, where 
$\{ \ ,\ \}_{\mbox{\tiny{Pb}}}$ denotes a Poisson bracket and 
$[\ , \ ]$ a commutator of operators on a Fock space. That is 
$$fg=gf \quad \mbox{but} \quad \hat{f}\hat{g} \ne \hat{g}\hat{f}. $$
This describes the noncommutativity in Quantum Physics. 
Therefore there exist various operator orderings and we need to select a 
suitable one depending on a problem in question. 

Let $\adag $ and $a$ be the creation operator and the annihilation 
operator of a harmonic oscillator respectively. The relation is 
$$ [a,\ \adag ]=1 .$$
We put the number operator $ \ N \equiv \adag a \ $. 

There are various useful orderings in Quantum Physics \cite{CG}. 
In this paper, we deal with normal ordering,  
anti-normal ordering, and Weyl ordering. 
We denote them as follows: 
\begin{enumerate}
 \item Normal Ordering \qquad $:(\adag )^n a^m:_N \equiv (\adag)^n a^m$, 
 \item Anti-Normal Ordering \qquad 
$:(\adag )^n a^m:_{AN} \equiv a^m (\adag)^n$, 
 \item Weyl Ordering \\
$:(\adag)^{n} a^m:_W \equiv 
 {\begin{pmatrix}
 n+m \\
 n
 \end{pmatrix}}^{-1}
( \mbox{sum of all symmetric products of $n \adag$ and $m a$} )$, 
\end{enumerate}
For example, when $m=n$, 
\begin{eqnarray}
:\adag a:_W &=& \displaystyle \frac 12 (\adag a+a\adag ), \label{eqn:W1}\\
&& \nonumber\\
:(\adag a)^2:_W &=& \displaystyle \frac 16 
 (\adag \adag aa+\adag a\adag a+\adag aa\adag 
 +a\adag \adag a+a\adag a\adag +aa\adag \adag ), \label{eqn:W2}\\
&& \nonumber\\
:(\adag a)^3:_W &=& \displaystyle \frac{1}{20} 
 (\adag \adag \adag aaa+\adag \adag a\adag aa
 +\adag \adag aa\adag a+\adag \adag aaa\adag + \nonumber\\ && \hspace{5mm}
 \adag a\adag \adag aa+\adag a\adag a\adag a
 +\adag a\adag aa\adag +\adag aa\adag \adag a+ \nonumber\\ && \hspace{5mm}
 \adag aa\adag a\adag +\adag aaa\adag \adag 
 +a\adag \adag \adag aa+a\adag \adag a\adag a+ \nonumber\\ && \hspace{5mm}
 a\adag \adag aa\adag +a\adag a\adag \adag a
 +a\adag a\adag a\adag +a\adag aa\adag \adag + \nonumber\\ && \hspace{5mm}
 aa\adag \adag \adag a+aa\adag \adag a\adag 
 +aa\adag a\adag \adag +aaa\adag \adag \adag ), \label{eqn:W3}
\end{eqnarray}
here we write $:(\adag )^n a^n :_{*}$ as $:(\adag a)^n:_{*}$ for simplicity 
($*$=N, AN, W). 

We often make proper use of these orderings. Therefore it is important to 
investigate what are the relations among them. 
The intertwining formulas of these orderings have been known by 
\cite{CG1}, \cite{CG2} as 
a formula of ``$s$-ordered power-series expansions";
\begin{equation}
\{ (\adag)^n a^m \}_s=
 \sum_{k=0}^{\min{\{n, m \} }} 
 k! \ 
 {\begin{pmatrix} n \\ k \end{pmatrix}}
 {\begin{pmatrix} m \\ k \end{pmatrix}}
 \left( \frac{t-s}{2} \right)^k \ 
 \{ (\adag )^{n-k}a^{m-k} \}_t, 
  \label{eqn:s-order}
\end{equation}
where
$$
\{ (\adag)^n a^m \}_s \equiv 
\left.
\frac{\partial^{n+m}}{\partial \alpha^n \partial (-\alpha^{*})^m}
\mbox{e}^{\alpha \adag - \alpha^{*} a}\mbox{e}^{s|\alpha|^2/2}
\right|_{\alpha=0}, 
$$
and the ordering specified $s,t=+1, 0, -1$ are, respectively, 
normal, Weyl, and anti-normal. 
On the other hand, the intertwining formulas among three orderings are 
used in the Weyl calculus. Recently, Omori et al. \cite{OMMY}
have investigated strange phenomena of $*$-exponential functions 
of quadratic forms in the Weyl algebra. 
They have written the intertwining formulas as a form of linear operators 
by using the product formulas of the $*$-exponential functions; 
\begin{equation}\label{eqn:N-W}
 \displaystyle \mbox{e}^{\frac 12 \partial_{\adag}\partial_a}
 :f(\adag, \ a):_N
 \ = \ :f(\adag, \ a):_W, 
\end{equation}
\begin{equation}\label{eqn:AN-W}
 \displaystyle \mbox{e}^{-\frac 12 \partial_{\adag}\partial_a}
 :f(\adag, \ a):_{AN}
 \ = \ :f(\adag, \ a):_W, 
\end{equation}
where $f(\adag, a)$ is a suitable 2-variable function $f(\bar{z}, z)$ 
substituting $\adag, a$ for $\bar{z}, z$ respectively. 

Under these circumstances, 
we investigate expressions of Weyl ordering 
with the intertwining formulas of orderings. The motivation is 
as follows; The definition of 
Weyl ordering is clear, but it is hard to give an explicit expression of it, 
like (\ref{eqn:W1}), (\ref{eqn:W2}), (\ref{eqn:W3}), 
especially when $n$ is large. 
Therefore it is important to search a simple expression of it. 
If we study by using the formula (\ref{eqn:s-order}) or (\ref{eqn:N-W}), 
(\ref{eqn:AN-W}), 
though (\ref{eqn:s-order}) give some explicit expressions of Weyl ordering, 
it is not ``symmetric" that Weyl ordering would be. 
Therefore, we study this problem in another point of view, 
namely in the combinatorial theory and 
in the difference theory. Then we find a symmetric expression of 
Weyl ordering by using the number operator $N$, which is new and 
would be useful in Quantum Physics. 
Since $\adag a=N$ and $a\adag =N+1$, we shall show Weyl ordering 
$:(\adag a)^n:_W$ as a symmetric polynomial with respect to $N$ and $N+1$.

For example, (\ref{eqn:W1}), (\ref{eqn:W2}), (\ref{eqn:W3}) are 
\begin{eqnarray}
:\adag a:_W &=& \displaystyle \frac 12 \{ N+(N+1) \} , \label{eqn:Ws1}\\
&& \nonumber\\
:(\adag a)^2:_W &=& \displaystyle \frac 12 \{ N^2+(N+1)^2 \} , 
\label{eqn:Ws2}\\
&& \nonumber\\
:(\adag a)^3:_W &=& \displaystyle \frac 12 \{ N^3+(N+1)^3 \} +
 \frac 14 \{ N+(N+1) \} . \label{eqn:Ws3}
\end{eqnarray}

In fact, for all $n=1,2,\cdots,$ we can show $:(\adag a)^n:_W$ 
as a symmetric form like these examples in section 6. 
It is a new and very clear-cut expression of Weyl ordering. 

The contents of this paper are as follows. 
In section 2, we prepare the difference theory and 
the Stirling number of the first kind. 
In section 3, first we prepare some facts about normal 
and anti-normal ordering. Next, we give a simple proof 
for the intertwining formulas of orderings as above for a help of 
the following sections. 
In section 4, we interpret these expressions by 
the difference theory. Then we find that the noncommutativity of 
harmonic oscillators is nothing but 
the the Newton expansion w.r.t. variable $N$. 
In section 5,  we describe nontrivial relations including 
the Stirling number of the first kind. 
These relations are derived from the noncommutativity of 
harmonic oscillators and are needed in the following section. 
In section 6, we describe the main result, that is a new symmetric 
expression of Weyl ordering. 
In section 7, we summarize our results. 

\section{Mathematical Preliminaries}
In this section, we prepare the difference theory and 
the Stirling number of the first kind. 

First we define factorial of degree $n$ with an increment $\epsilon$ \cite{J}
\begin{equation}
\xnU \equiv x(x-\epsilon) \cdots (x-(n-1)\epsilon)
\end{equation}
and 
\begin{equation}
x^{\overline{n}} \equiv x(x+\epsilon) \cdots (x+(n-1)\epsilon). 
\end{equation}
We remark 
\begin{equation}
 \xnO=(x+\epsilon) \cdots (x+n\epsilon)=(x+n\epsilon)^{\underline{n}}. 
\end{equation}
Let $p_n(x)$ be a polynomial of degree $n$ w.r.t. $x$, and $\Delta_{+x}$ 
the forward difference operator defined by 
$\Delta_{+x}f(x) \equiv \epsilon^{-1} \{ f(x+\epsilon)-f(x) \} . $
Then we have the Newton expansion
\begin{equation}
 p_n(x)=\sum_{k=0}^n \frac{x^{\underline{k}}}{k!} \Delta_{+x}^k p_n(0) 
  = \sum_{k=0}^n \frac{x^{\underline{n-k}}}{(n-k)!} \Delta_{+x}^{n-k} p_n(0).
\label{eqn:expand}
\end{equation}
We also remark that 
\begin{equation}
 \Delta_{+x} x^{\underline{n}}=n x^{\underline{n-1}}. 
\end{equation}

Next we define the Stirling number of the first kind and 
calculate some of them explicitly. 
\begin{df}The Stirling number of the first kind 
$\ s(n,i) \quad (i=1, \cdots, n) \ $ 
is defined by the following equation \cite{R}; 
$$ \xnU=x(x-\epsilon) \cdots (x-(n-1)\epsilon)
=\sum_{i=1}^n s(n,i)\epsilon^{n-i} x^i $$
and $\ s(n,0) \equiv 0 \quad (n \geq 1) \ $, 
$\ s(j,i) \equiv 0 \quad (j<i) \ $. 
\end{df}
\noindent
We remark  
\begin{equation}\label{eqn:def1}
 x^{\overline{n}}=x(x+\epsilon) \cdots (x+(n-1)\epsilon)
 =\sum_{i=1}^n (-1)^{n-i} s(n,i) \epsilon^{n-i} x^i. 
\end{equation}

\begin{rmk}
We can obtain the Stirling number of the first kind explicitly 
by the following recursion formula and initial conditions. 
\begin{eqnarray}
s(n+1,i) &=& s(n,i-1)-n \cdot s(n,i) \quad (n \geq i \geq 1), 
 \label{eqn:recur}\\
s(n,n) &=& 1 \quad (n \geq 1), \label{eqn:Sn-0}\\
s(n,0) &=& 0. 
\end{eqnarray}
\end{rmk}
\noindent
For example, if we put $i=n$ in (\ref{eqn:recur}), then 
\begin{eqnarray*}
s(n+1,n) &=& s(n,n-1)-n \cdot s(n,n) \\
 &=& s(n,n-1)-n. 
\end{eqnarray*}
Therefore 
\begin{equation}\label{eqn:Sn-1}
 s(n,n-1)=-\sum_{k=1}^{n-1}k=-\frac{n(n-1)}{2}. 
\end{equation}
Next, since 
\begin{eqnarray*}
s(n+1,n-1) &=& s(n,n-2)-n \cdot s(n,n-1) \\
 &=& s(n,n-2)+\frac 12 n^2(n-1), 
\end{eqnarray*}
we have 
\begin{equation}\label{eqn:Sn-2}
 s(n,n-2)=\sum_{k=1}^{n-1}\frac 12 k^2(k-1)
 =\frac{1}{24}n(n-1)(n-2)(3n-1). 
\end{equation}

\section{A Simple Proof for the Intertwining Formulas of Orderings}

First we prepare some facts about normal and anti-normal ordering. 
For a while, we put $[a,\ \adag ]=\epsilon$ for more general situation. 
\begin{lem}\label{lem:2-1}
We have following formulas using the number operator $N=\adag a $; 
\begin{equation}\label{eqn:nor}
:(\adag a)^n:_N =N(N-\epsilon) \cdots (N-(n-1)\epsilon)=\NnU,
\end{equation}
\begin{equation}\label{eqn:antinor}
:(\adag a)^n:_{AN} =(N+\epsilon)(N+2\epsilon) \cdots (N+n\epsilon)=\NnO.
\end{equation}
\end{lem}

\begin{proof}
By using simple relations 
$$ aN=(N+\epsilon)a, \ \adag N=(N-\epsilon)\adag \quad \mbox{or} 
\quad Na=a(N-\epsilon), \ N\adag =\adag (N+\epsilon), $$
we have an induction foumula
\begin{eqnarray*}
 (\adag)^n a^n &=& \adag \cdots \adag \adag \adag a aa \cdots a \\
  &=& \adag \cdots \adag \adag N aa \cdots a \\
  &=& \adag \cdots \adag (N-\epsilon) \adag aa \cdots a \\
  &=& \cdots \\
  &=& (N-(n-1)\epsilon) (\adag)^{n-1} a^{n-1}. 
\end{eqnarray*}
Therefore we obtain (\ref{eqn:nor}) by the mathematical induction. 
We also obtain (\ref{eqn:antinor}) in a similar way. 
\end{proof}

As we mentioned in section $1$, since Omori et al have proved 
(\ref{eqn:N-W}), (\ref{eqn:AN-W}) by using the product formulas of 
the $*$-exponential functions, 
we give an another simple proof of them. We have only to prove the next 
proposition. 
This proof helps account for our theory to be described in the 
following sections.

\begin{prop}\label{prop:intertwiner}
\begin{equation}
 \displaystyle \mbox{e}^{\frac 12 \epsilon \partial_{\adag}\partial_a}
  :(\adag)^n a^m:_N
 \ = \ :(\adag)^n a^m:_W, \label{eqn:formula-nor}
\end{equation}
\begin{equation}
 \displaystyle \mbox{e}^{-\frac 12 \epsilon \partial_{\adag}\partial_a}
  :(\adag)^n a^m:_{AN}
 \ = \ :(\adag)^n a^m:_W. \label{eqn:formula-antinor}
\end{equation}
\end{prop}
\begin{proof}
By using the generating function of the Weyl ordering, 
$$ \mbox{e}^{\alpha \adag +\beta a}=\sum_{n,m=0}^{\infty}
 \frac{\alpha^n \beta^m}{n!m!}:(\adag)^n a^m:_W $$
and the fundamental Baker-Campbell-Hausdorff formula \cite{FS}
\begin{equation}
 \mbox{e}^{A+B}=\mbox{e}^{-\frac12[A,B]}\mbox{e}^{A}\mbox{e}^{B}
  \quad \mbox{whenever $([A,[A,B]]=[B,[A,B]]=0$), }
   \label{eqn:BKH}
\end{equation}
we have \cite{KOS}
\begin{eqnarray}
\mbox{e}^{\alpha \adag +\beta a}
&=& \mbox{e}^{\frac{1}{2}\epsilon \alpha \beta}
    \mbox{e}^{\alpha \adag}\mbox{e}^{\beta a} \qquad 
     (\mbox{normal order}) \nonumber\\
&=& \mbox{e}^{\frac 12 \epsilon \partial_{\adag}\partial_a}
    \mbox{e}^{\alpha \adag}\mbox{e}^{\beta a}. \label{eqn:intertwiner1}
\end{eqnarray}
If we expand both sides of (\ref{eqn:intertwiner1}) and compare 
the coefficients of $\alpha^n \beta^m$, we obtain (\ref{eqn:formula-nor}). 
Similarly, we have 
\begin{eqnarray}
\mbox{e}^{\alpha \adag +\beta a}
&=& \mbox{e}^{-\frac{1}{2}\epsilon \alpha \beta}
    \mbox{e}^{\beta a}\mbox{e}^{\alpha \adag} \qquad 
     (\mbox{anti-normal order}) \nonumber\\
&=& \mbox{e}^{-\frac 12 \epsilon \partial_{\adag}\partial_a}
    \mbox{e}^{\alpha \adag}\mbox{e}^{\beta a}. \label{eqn:intertwiner2}
\end{eqnarray}
If we expand both sides of (\ref{eqn:intertwiner2}) and compare 
the coefficients of $\alpha^n \beta^m$, 
we obtain (\ref{eqn:formula-antinor}).
\end{proof}

By the proposition \ref{prop:intertwiner} and the lemma \ref{lem:2-1}, 
we obtain the following corollary. 

\begin{cor}\label{cor:3-3}
 We can show Weyl ordering $:(\adag)^{n} a^m:_W$ in terms of 
normal ordering. 
\begin{equation}
:(\adag)^n a^m:_W=
 \sum_{k=0}^{\min{\{n, m \} }} 
 \frac{\epsilon^k k!}{2^k} \ 
 {\begin{pmatrix} n \\ k \end{pmatrix}}
 {\begin{pmatrix} m \\ k \end{pmatrix}}
 \ (\adag )^{n-k}a^{m-k}. 
  \label{eqn:st=01}
\end{equation}
\noindent
Especially, in case of $m=n$, if we use (\ref{eqn:nor}), 
we obtain an expression of $:(\adag a)^n:_W$ in terms of 
the number operator $N$, 
\begin{equation}\label{eqn:Wnor}
:(\adag a)^n:_W=
 \sum_{k=0}^n \frac{\epsilon^k k!}{2^k} \ 
 {\begin{pmatrix} n \\ k \end{pmatrix}}^2 \NnkU. 
\end{equation}
\end{cor}

By a similar calculation, we obtain the next expression using 
anti-normal ordering. 

\begin{cor}\label{cor:3-4}
\begin{equation}
:(\adag)^n a^m:_W=
 \sum_{k=0}^{\min{\{n, m \} }} 
 \frac{(-\epsilon)^k k!}{2^k} \ 
 {\begin{pmatrix} n \\ k \end{pmatrix}}
 {\begin{pmatrix} m \\ k \end{pmatrix}}
 \ a^{n-k}(\adag )^{m-k}. 
  \label{eqn:st=0-1}
\end{equation}
\noindent
Especially, in case of $m=n$, if we use (\ref{eqn:antinor}), 
we obtain an another expression of $:(\adag a)^n:_W$ 
in terms of the number operator $N$, 
\begin{equation}\label{eqn:Wantinor}
:(\adag a)^n:_W=
 \sum_{k=0}^n \frac{(-\epsilon)^k k!}{2^k} \ 
 {\begin{pmatrix} n \\ k \end{pmatrix}}^2 \NnkO. 
\end{equation}
\end{cor}

\begin{rmk}
In \cite{CG1}, \cite{CG2}, K.E.Cahill and R.J.Glauber introduced 
``$s$-ordered power-series expansions" as follows; \\
Define the $s$-ordered displacement operator $D(\alpha, s)$ by 
\begin{equation}
D(\alpha, s)=\mbox{e}^{\alpha \adag-\alpha^{*} a}
 \mbox{e}^{s|\alpha|^2/2}, \qquad \mbox{where $\alpha \in \mathbf{C}$
 and $\alpha^{*}$ is its complex conjugate.}
  \label{eqn:displacement}
\end{equation}
By the fundamental Baker-Campbell-Hausdorff formula (\ref{eqn:BKH}), 
for the three discrete values of $s=+1, 0, -1$, the operator 
$D(\alpha, s)$ can be written as an exponential which is, respectively, 
normal order 
$$
D(\alpha, 1)=\mbox{e}^{\alpha \adag}\mbox{e}^{-\alpha^{*} a}, 
$$
Weyl order 
$$
D(\alpha, 0)=\mbox{e}^{\alpha \adag-\alpha^{*} a}, 
$$
and anti-normal order 
$$
D(\alpha, -1)=\mbox{e}^{-\alpha^{*} a}\mbox{e}^{\alpha \adag}. 
$$
They defined the $s$-ordered product $\{ (\adag)^n a^m \}_s$ by means of 
the Taylor series 
$$
D(\alpha, s)=\sum_{n,m=0}^{\infty} \{ (\adag)^n a^m \}_s 
 \frac{{\alpha}^n (-\alpha^{*})^m}{n!m!}
$$
or, equivalently, 
$$
\{ (\adag)^n a^m \}_s \equiv 
\left.
\frac{\partial^{n+m} D(\alpha, s)}
{\partial \alpha^n \partial (-\alpha^{*})^m}
\right|_{\alpha=0}. 
$$
The intertwining formula between $s$-ordered product and 
$t$-ordered product is 
$$
D(\alpha, s)=\mbox{e}^{(s-t)|\alpha|^2/2} D(\alpha, t)
$$
or 
\begin{equation}
\{ (\adag)^n a^m \}_s=
 \sum_{k=0}^{\min{\{n, m \} }} 
 k! \ 
 {\begin{pmatrix} n \\ k \end{pmatrix}}
 {\begin{pmatrix} m \\ k \end{pmatrix}}
 \left( \frac{t-s}{2} \right)^k \ 
 \{ (\adag )^{n-k}a^{m-k} \}_t
  \label{eqn:s-order1}. 
\end{equation}
If we put $(s,t)=(0,1), (0,-1)$, we recover (\ref{eqn:st=01}), 
(\ref{eqn:st=0-1}) in case of $\epsilon=1$ respectively. 

We remark that since they did not write 
in the form of (\ref{eqn:formula-nor}) and (\ref{eqn:formula-antinor})
in \cite{CG1}, \cite{CG2} explicitly, though the idea of the proof was 
almost the same, we gave the proof of them in this section. 
\end{rmk}

\section{A Relation Between the Noncommutativity of Harmonic Oscillators 
and the Difference Theory}

In this section, we describe a relation between the noncommutativity of 
harmonic oscillators and the difference theory. 
The noncommutativity of harmonic oscillators is as follows; 
\begin{equation}
\mbox{e}^{t\alpha \beta}\mbox{e}^{\alpha \adag}\mbox{e}^{\beta a}
=\mbox{e}^{s\alpha \beta}\mbox{e}^{\beta a}\mbox{e}^{\alpha \adag}, 
  \label{eqn:noncom1}
\end{equation}
where $t-s=\epsilon$. 
If we expand both sides of (\ref{eqn:noncom1}), then we have 
\begin{equation}
\displaystyle \sum_{k,l,m}\frac{t^k \alpha^k \beta^k}{k!}
 \frac{\alpha^l (\adag)^l}{l!}\frac{\beta^m a^m}{m!}
=\sum_{k,l,m}\frac{s^k \alpha^k \beta^k}{k!}
 \frac{\beta^l a^l}{l!}\frac{\alpha^m (\adag)^m}{m!}.
  \label{eqn:noncom2}
\end{equation}
We choose coefficients of $\alpha^n \beta^n$ of (\ref{eqn:noncom2}), 
then the left hand side is 
\begin{eqnarray*}
\displaystyle \sum_{k=0}^n \frac{t^k}{k!}
 \frac{(\adag)^{n-k}}{(n-k)!}\frac{a^{n-k}}{(n-k)!} 
&=& \displaystyle \frac{1}{(n!)^2} \sum_{k=0}^n k! \ t^k 
 {\begin{pmatrix}
 n \\
 k
 \end{pmatrix}}^2
 (\adag)^{n-k}a^{n-k} \\
&=& \displaystyle \frac{1}{(n!)^2} \sum_{k=0}^n k! \ t^k 
 {\begin{pmatrix}
 n \\
 k
 \end{pmatrix}}^2 \NnkU. 
\end{eqnarray*}
By a similar calculation of the right hand side, we obtain 
\begin{equation}
\sum_{k=0}^n k! \ t^k 
 {\begin{pmatrix}
 n \\
 k
 \end{pmatrix}}^2 \NnkU
=\sum_{k=0}^n k! \ s^k 
 {\begin{pmatrix}
 n \\
 k
 \end{pmatrix}}^2 \NnkO. \label{eqn:noncom3}
\end{equation}
\begin{rmk}
In the case of 
$\displaystyle t=\frac{\epsilon}{2}$ and 
$\displaystyle s=-\frac{\epsilon}{2}$, 
(\ref{eqn:noncom3}) is the equation that 
(\ref{eqn:Wnor}) and (\ref{eqn:Wantinor}) are equal. 
Moreover, since $(\adag)^n a^n=N^{\underline{n}}$ and 
\begin{equation}
 \partial_{\adag} (\adag)^n a^n= n(\adag)^{n-1} a^n=
 nN^{\underline{n-1}} \ a, \qquad  
 \Delta_+ N^{\underline{n}}=nN^{\underline{n-1}}, \qquad 
 (\Delta_+ =\Delta_{+N})
\end{equation}
relations using various orderings of 
$(\adag)^{n-1} a^n=\displaystyle \frac{\partial_{\adag}}{n}(\adag)^n a^n$
are equivalent to the relation (\ref{eqn:noncom3}) acted by $\Delta_+$ 
with some $t$. 
Similarly, for any $n,\ m$, relations using various orderings of 
$(\adag)^n a^m$ are equivalent to the relation 
(\ref{eqn:noncom3}) acted by some power of $\Delta_+$ and some $t$. 
\end{rmk}

\begin{rmk}From (\ref{eqn:noncom3}), 
$$ \displaystyle \frac{d}{dt} \sum_{k=0}^n k! \ t^k 
 {\begin{pmatrix} n \\ k \end{pmatrix}}^2 \NnkU
=\displaystyle \frac{d}{dt} \sum_{k=0}^n k! \ (t-\epsilon)^k 
 {\begin{pmatrix} n \\ k \end{pmatrix}}^2 \NnkO $$
$$ \sum_{k=0}^n k! \cdot k \ t^{k-1} 
 {\begin{pmatrix} n \\ k \end{pmatrix}}^2 \NnkU
= \sum_{k=0}^n k! \cdot k \ (t-\epsilon)^{k-1} 
 {\begin{pmatrix} n \\ k \end{pmatrix}}^2 \NnkO . $$
Therefore, if we put 
$\displaystyle t=\frac 12 $ and $\epsilon=1$, 
we obtain the following equation 
\begin{equation}
\sum_{k=0}^n \displaystyle  \frac{k! \cdot k}{2^{k-1}} \ 
 {\begin{pmatrix} n \\ k \end{pmatrix}}^2 \ 
 \left\{ \NnkU +(-1)^k (N+1)^{\overline{n-k}} \right\} =0. 
  \label{eqn:n-1}
\end{equation}
\end{rmk}

Next we show the equation of noncommutativity of harmonic oscillators 
(\ref{eqn:noncom3}) by using the difference theory. 
We put
\begin{equation}
 p_n(N)=\sum_{k=0}^n k! \ s^k 
 {\begin{pmatrix}
 n \\
 k
 \end{pmatrix}}^2 \NnkO
=\sum_{k=0}^n k! \ s^k 
 {\begin{pmatrix}
 n \\
 k
 \end{pmatrix}}^2 (N+(n-k)\epsilon)^{\underline{n-k}}, 
\label{eqn:noncom4}
\end{equation}
and we expand (\ref{eqn:noncom4}) using the forward difference operator 
$\Delta_+$. 
Since 
\begin{eqnarray*}
\Delta_+^{n-l} p_n(0) &=& \sum_{k=0}^n k! \ s^k 
 {\begin{pmatrix}
 n \\
 k
 \end{pmatrix}}^2 (n-k)(n-k-1) \cdots (l-k+1) \ 
 (0+(n-k)\epsilon)^{\underline{l-k}} \\
&=& \sum_{k=0}^n k! \ s^k 
 {\begin{pmatrix}
 n \\
 k
 \end{pmatrix}} \frac{n!}{k!(n-k)!} \ (n-k)(n-k-1) \cdots (l-k+1) \ 
 \frac{(n-k)!}{(n-l)!} \epsilon^{l-k} \\
&=& l! 
 {\begin{pmatrix}
 n \\
 l
 \end{pmatrix}} \sum_{k=0}^n \ s^k 
 {\begin{pmatrix}
 n \\
 k
 \end{pmatrix}} \left( \frac{d}{d\epsilon} \right)^{n-l} 
  \epsilon^{n-k} \\
&=& l! 
 {\begin{pmatrix}
 n \\
 l
 \end{pmatrix}} \left( \frac{d}{d\epsilon} \right)^{n-l} 
 (s+\epsilon)^n \\
&=& l! 
 {\begin{pmatrix}
 n \\
 l
 \end{pmatrix}} \ \frac{n!}{l!} t^l, 
\end{eqnarray*}
by (\ref{eqn:expand}), we have 
\begin{eqnarray*}
p_n(N) &=& \sum_{l=0}^n \Delta_+^{n-l} p_n(0) 
 \frac{N^{\underline{n-l}}}{(n-l)!} \\
&=& \sum_{l=0}^n l! 
 {\begin{pmatrix}
 n \\
 l
 \end{pmatrix}} \ \frac{n!}{l!} t^l \ 
 \frac{N^{\underline{n-l}}}{(n-l)!} \\
&=& \sum_{l=0}^n l! \ t^l 
 {\begin{pmatrix}
 n \\
 l
 \end{pmatrix}}^2 
 N^{\underline{n-l}}. 
\end{eqnarray*}
Therefore, the noncommutativity of harmonic oscillators is nothing but 
the the Newton expansion w.r.t. variable $N$. More precisely, 
the noncommutative parameter $\epsilon$ in $[a, \ \adag]=\epsilon$ 
corresponds to the increment of the difference operator 
$\Delta_+=\Delta_{+N}$. 

\section{Relations Including the Stirling Number of the First Kind}

We shall describe nontrivial relations including 
the Stirling number of the first kind. 
These relations are derived from the noncommutativity of 
harmonic oscillators. Hereafter, we put $\epsilon=1$ again. 

\begin{thm}\label{thm:relation}
We have 
\begin{equation}\label{eqn:rel}
\sum_{k=0}^{2j+1} 
 \frac{k!}{2^k} \ 
 {\begin{pmatrix} n-1 \\ k \end{pmatrix}}
 {\begin{pmatrix} n \\ k \end{pmatrix}} \ s(n-k,n-2j-1)=0 
  \quad (j=0,1,\cdots). 
\end{equation}
\end{thm}

\noindent
For example, 
\begin{eqnarray*}
\sum_{k=0}^{1} \frac{k!}{2^k} \ 
 {\begin{pmatrix} n-1 \\ k \end{pmatrix}}
 {\begin{pmatrix} n \\ k \end{pmatrix}} \ s(n-k,n-1) 
&=& s(n,n-1)+\displaystyle \frac{(n-1)n}{2} s(n-1,n-1) \\
&=& \displaystyle -\frac{n(n-1)}{2}+\displaystyle \frac{(n-1)n}{2} \\
&=& 0. 
\end{eqnarray*}

This theorem is used in a new symmetric expression 
of $:(\adag a)^n:_W$ with respect to $\adag a=N$ and $a\adag =N+1$. 

\begin{proof}From (\ref{eqn:noncom3}), 
$$ \sum_{k=0}^n k! \ t^k 
 {\begin{pmatrix} n \\ k \end{pmatrix}}^2 \Delta_+ \NnkU
= \sum_{k=0}^n k! \ (t-1)^k 
 {\begin{pmatrix} n \\ k \end{pmatrix}}^2 \Delta_+ (N+n-k)^{\underline{n-k}}$$
$$ \sum_{k=0}^n k! \ t^k 
 {\begin{pmatrix} n \\ k \end{pmatrix}}^2 (n-k) N^{\underline{n-k-1}}
= \sum_{k=0}^n k! \ (t-1)^k
 {\begin{pmatrix} n \\ k \end{pmatrix}}^2 (n-k) (N+n-k)^{\underline{n-k-1}}.$$
Therefore, if we put 
$\displaystyle t=\frac 12 $ , 
we obtain the following equation. 
\begin{equation}
\sum_{k=0}^n \displaystyle  \frac{k! \cdot (n-k)}{2^k} \ 
 {\begin{pmatrix} n \\ k \end{pmatrix}}^2 \ 
 \left\{ N^{\underline{n-k-1}} -(-1)^k (N+n-k)^{\underline{n-k-1}} \right\} 
 =0. \label{eqn:Delta}
\end{equation}
If we multiply (\ref{eqn:Delta}) by $(N+1)$, then \ 
$(N+n-k)^{\underline{n-k-1}} \cdot (N+1)=(N+1)^{\overline{n-k}}$ \ and 
the left hand side is 
\begin{eqnarray*}
&& \sum_{k \geq 0} \displaystyle  \frac{k!}{2^k} \ n 
 {\begin{pmatrix} n-1 \\ k \end{pmatrix}}
 {\begin{pmatrix} n \\ k \end{pmatrix}} \ 
 \left\{ (N+1)^{\underline{n-k}} 
  +(-1)^{k+1} (N+1)^{\overline{n-k}} \right\} \\
&=& n \sum_{k \geq 0} \displaystyle  \frac{k!}{2^k} \ 
 {\begin{pmatrix} n-1 \\ k \end{pmatrix}}
 {\begin{pmatrix} n \\ k \end{pmatrix}} \ 
 \left\{ \sum_{i=1}^{n-k} s(n-k,i) (N+1)^i 
  +(-1)^{k+1} \sum_{i=1}^{n-k} (-1)^{n-k-i} s(n-k,i) (N+1)^i \right\} \\
&=& n \sum_{i \geq 1} \sum_{k \geq 0} \displaystyle  \frac{k!}{2^k} \ 
 {\begin{pmatrix} n-1 \\ k \end{pmatrix}}
 {\begin{pmatrix} n \\ k \end{pmatrix}} \ s(n-k,i)
 \left\{ 1+(-1)^{n-1-i} \right\} (N+1)^i \\
&=& 2n \sum_{j \geq 0} \sum_{k \geq 0} \displaystyle  \frac{k!}{2^k} \ 
 {\begin{pmatrix} n-1 \\ k \end{pmatrix}}
 {\begin{pmatrix} n \\ k \end{pmatrix}} \ s(n-k,n-2j-1)
 (N+1)^{n-2j-1}. 
\end{eqnarray*}
Therefore we obtain the theorem \ref{thm:relation}. 
\end{proof}

\begin{rmk}
More generally, by the equation 
$$ \mbox{e}^{\frac 12 \partial_{\adag}\partial_a}:(\adag)^{n+m} a^n:_N
=:(\adag)^{n+m} a^n:_W
=\mbox{e}^{-\frac 12 \partial_{\adag}\partial_a}:(\adag)^{n+m} a^n:_{AN}, $$
we obtain the next theorem. 

\begin{thm}
For $n>0$ and $m=-n+1, -n+2, \cdots, -1, 0$, we have 
\begin{eqnarray}
&& \{ (-1)^{n+m-i}+1 \} \sum_{k=0}^{n+m+1-i} 
 \frac{k!}{2^k} \ 
 {\begin{pmatrix} n+m \\ k \end{pmatrix}}
 {\begin{pmatrix} n \\ k \end{pmatrix}}
 \ s(n+m+1-k,i) \nonumber\\
&+& \sum_{l=i+1}^{n+m+1} (m+1)^{l-i} \ 
 {\begin{pmatrix} l-1 \\ i-1 \end{pmatrix}} \ 
 \sum_{k=0}^{n+m+1-l} 
 \frac{k!}{2^k} \ 
 {\begin{pmatrix} n+m \\ k \end{pmatrix}}
 {\begin{pmatrix} n \\ k \end{pmatrix}}
 \ s(n+m+1-k,l)=0 \label{eqn:n-m-rel}\\
&& \hspace{8cm} (\mbox{for} \ i=1,2, \cdots, n+m+1), \nonumber
\end{eqnarray}
and for $n>0$ and $m=0, 1, 2, \cdots $, we have 
\begin{eqnarray}
&& \{ (-1)^{n-i}+1 \} \sum_{k=0}^{n+1-i} 
 \frac{k!}{2^k} \ 
 {\begin{pmatrix} n+m \\ k \end{pmatrix}}
 {\begin{pmatrix} n \\ k \end{pmatrix}}
 \ s(n+1-k,i) \nonumber\\
&+& \sum_{l=i+1}^{n+1} (1-m)^{l-i} \ 
 {\begin{pmatrix} l-1 \\ i-1 \end{pmatrix}} \ 
 \sum_{k=0}^{n+1-l} 
 \frac{k!}{2^k} \ 
 {\begin{pmatrix} n+m \\ k \end{pmatrix}}
 {\begin{pmatrix} n \\ k \end{pmatrix}}
 \ s(n+1-k,l)=0 \label{eqn:n-m-rel2}\\
&& \hspace{8cm} (\mbox{for} \ i=1,2, \cdots, n+1). \nonumber
\end{eqnarray}
\end{thm}
\end{rmk}

\section{A Symmetric Expression of Weyl Ordering}

We have expressed $:(\adag a)^n:_W$ as a polynomial of the number operator 
$N$ in two ways, namely normal ordering expression (\ref{eqn:Wnor}) 
and anti-normal one (\ref{eqn:Wantinor}). 
Moreover, we have a new symmetric expression of $:(\adag a)^n:_W$ 
with respect to $\adag a=N$ and $a\adag =N+1$. 

\begin{thm}\label{thm:WsymmPoly}
\begin{equation}\label{eqn:Wsymm}
:(\adag a)^n:_W=\sum_{j=0}^{[(n-1)/2]} \alpha (n,2j) 
 \{ N^{n-2j}+(N+1)^{n-2j} \} , 
\end{equation}
where the constants $\alpha (n,i)$ are defined by 
$$ \alpha (n,i) \equiv \frac 12 \ 
\sum_{k=0}^{i} \frac{k!}{2^k} \ 
{\begin{pmatrix} n \\ k \end{pmatrix}}
{\begin{pmatrix} n-1 \\ k \end{pmatrix}}
\ s(n-k,n-i) 
\qquad (i=1,2,\cdots), $$ and we have 
$\alpha (n,2j+1)=0 \ (j=0,1,\cdots)$. 
\end{thm}
\noindent
For example, 
\begin{equation}\label{eqn:alphan0}
\alpha (n,0) = \frac 12, 
\end{equation}
and by using (\ref{eqn:Sn-0}), (\ref{eqn:Sn-1}) and (\ref{eqn:Sn-2}), 
we have easily
\begin{eqnarray}
\alpha (n,2) &=& \frac 12 \ 
\sum_{k=0}^{2} \frac{k!}{2^k} \ 
{\begin{pmatrix} n \\ k \end{pmatrix}}
{\begin{pmatrix} n-1 \\ k \end{pmatrix}}
\ s(n-k,n-2) 
\nonumber\\
&=& \frac 12
\left\{ s(n,n-2)+\frac 12 n(n-1) \cdot s(n-1,n-2)
+\frac{1}{8} n(n-1)^2(n-2) \cdot s(n-2,n-2) \right\} 
\nonumber\\
&=& \frac{1}{24}n(n-1)(n-2). \label{eqn:alphan2}
\end{eqnarray}
We remark that if we put $n=1,2,3$ in 
(\ref{eqn:alphan0}), (\ref{eqn:alphan2}), we recover 
(\ref{eqn:Ws1}), (\ref{eqn:Ws2}) and (\ref{eqn:Ws3}). 

By a similar calculation, we have 
\begin{eqnarray*}
&& \alpha (n,4)=\frac{1}{2880}n(n-1)(n-2)(n-3)(n-4)(5n-7), \\
&& \\
&& \alpha (n,6)=\frac{1}{725760}n(n-1)(n-2)(n-3)(n-4)(n-5)(n-6)
 (35n^2-147n+124), 
\end{eqnarray*}
where we used Mathematica. 


\begin{proof}
First, by (\ref{eqn:n-1}), we remark that we have 
\begin{eqnarray*}
&& \sum_{k=0}^n \displaystyle \frac{k!}{2^k} \ 
 {\begin{pmatrix} n \\ k \end{pmatrix}}
 {\begin{pmatrix} n-1 \\ k-1 \end{pmatrix}} \ 
 \left\{ \NnkU +(-1)^k (N+1)^{\overline{n-k}} \right\} \\
&=& \frac{1}{2n}\sum_{k=0}^n \displaystyle  \frac{k! \cdot k}{2^{k-1}} \ 
 {\begin{pmatrix} n \\ k \end{pmatrix}}^2 \ 
 \left\{ \NnkU +(-1)^k (N+1)^{\overline{n-k}} \right\} \\
&=& 0. 
\end{eqnarray*}
By the proposition \ref{prop:intertwiner} and 
the corollary \ref{cor:3-3}, \ref{cor:3-4}, 
\begin{eqnarray*}
:(\adag a)^n:_W &=&
 \displaystyle \frac 12 \left( 
 \mbox{e}^{\frac 12 \partial_{\adag}\partial_a}
  :(\adag a)^n:_N
+\mbox{e}^{-\frac 12 \partial_{\adag}\partial_a}
  :(\adag a)^n:_{AN} \right) \\
&=& \displaystyle \frac 12 \left\{ 
 \sum_{k=0}^n \frac{k!}{2^k} \ 
 {\begin{pmatrix} n \\ k \end{pmatrix}}^2 \NnkU
+\sum_{k=0}^n \frac{(-1)^k k!}{2^k} \ 
 {\begin{pmatrix} n \\ k \end{pmatrix}}^2 (N+1)^{\overline{n-k}} \right\} \\
&=& \displaystyle \frac 12
 \sum_{k=0}^n \displaystyle \frac{k!}{2^k} \ 
 {\begin{pmatrix} n \\ k \end{pmatrix}}
 \left\{ 
 {\begin{pmatrix} n-1 \\ k \end{pmatrix}}
 +{\begin{pmatrix} n-1 \\ k-1 \end{pmatrix}} \right\} \ 
 \left\{ \NnkU +(-1)^k (N+1)^{\overline{n-k}} \right\} \\
&=& \displaystyle \frac 12
 \sum_{k=0}^{n-1} \displaystyle \frac{k!}{2^k} \ 
 {\begin{pmatrix} n \\ k \end{pmatrix}}
 {\begin{pmatrix} n-1 \\ k \end{pmatrix}} \ 
 \left\{ \NnkU +(-1)^k (N+1)^{\overline{n-k}} \right\} \\
&=& \displaystyle \frac 12
 \sum_{k=0}^{n-1} \displaystyle \frac{k!}{2^k} \ 
 {\begin{pmatrix} n \\ k \end{pmatrix}}
 {\begin{pmatrix} n-1 \\ k \end{pmatrix}} \ 
 \left\{ \sum_{i \geq 1} s(n-k,i) N^i \right. \\
&& \hspace{3cm} \left. 
 +(-1)^k \sum_{i \geq 1} (-1)^{n-k-i} s(n-k,i) (N+1)^i \right\} \\
&=& \sum_{i \geq 1} \left\{ 
\displaystyle \frac 12
 \sum_{k=0}^{n-i} \displaystyle \frac{k!}{2^k} \ 
 {\begin{pmatrix} n \\ k \end{pmatrix}}
 {\begin{pmatrix} n-1 \\ k \end{pmatrix}} \ 
  s(n-k,i) N^i \right. \\
&& \hspace{1cm} \left. 
+(-1)^{n-i} \displaystyle \frac 12
 \sum_{k=0}^{n-i} \displaystyle \frac{k!}{2^k} \ 
 {\begin{pmatrix} n \\ k \end{pmatrix}}
 {\begin{pmatrix} n-1 \\ k \end{pmatrix}} \
  s(n-k,i) (N+1)^i \right\} \\
&=& \sum_{i \geq 1} \alpha (n,n-i) \{ N^{i}+(-1)^{n-i} (N+1)^{i} \} , 
\end{eqnarray*}
where we used 
$s(n-k,i)=0 \quad (n-k<i)$. \\
If $i=n-2j-1 \quad (j=0,1,\cdots)$, $\alpha (n,2j+1)$ 
vanish because of (\ref{eqn:rel}). \\
And if $i=n-2j \quad (j=0,1,\cdots)$, 
since $(-1)^{n-i}=1$, we obtain the theorem \ref{thm:WsymmPoly}. 
\end{proof}

\section{Discussion}
We obtained a new symmetric expression of $:(\adag a)^n:_W$ 
w.r.t. $\adag a=N$ and $a\adag =N+1$ by using the difference theory. 
We also found that the noncommutative parameter 
$\epsilon$ 
corresponded to the increment of the difference operator w.r.t. variable $N$. 
Therefore, quantum (noncommutative) calculations of harmonic oscillators 
were done by classical (commutative) ones of the number operator 
by using the difference theory. 
As a by-product, nontrivial relations including the Stirling number 
of the first kind (\ref{eqn:n-m-rel}), (\ref{eqn:n-m-rel2}) 
were also obtained. 

\section*{Acknowledgements}
We are very grateful to Shin'ichi Nojiri for valuable discussions. 


\end{document}